\documentclass[twocolumn,amsmath,amssymb]{revtex4}
\usepackage{subfigure}
\usepackage{here}
\usepackage{float}
\usepackage{wrapfig}
\usepackage{graphicx}
\usepackage{dcolumn}
\usepackage{bm}
\usepackage{amsmath}
\usepackage{amsfonts}
\usepackage{amssymb}
\usepackage{color}
\usepackage{mathrsfs}
\makeatletter

\begin{document}
\title{Twist-induced snapping in a bent elastic ribbon}
\author{Tomohiko G. Sano$^{1,2}$}
\author{Hirofumi Wada$^{1}$} 

\affiliation{$^{1}$Department of Physical Sciences, Ritsumeikan University, Kusatsu, Shiga 525-8577, Japan}
\affiliation{$^{2}$Research Organization of Science and Technology, Ritsumeikan University, Kusatsu, Shiga 525-8577, Japan}

\begin{abstract}
Snapping of a slender structure is utilized in a wide range of natural and man-made systems, mostly to achieve rapid movement without relying on muscle-like elements. Although several mechanisms for elastic energy storage and rapid release have been studied in detail, a general understanding of the approach to design such a kinetic system is a key challenge in mechanics. Here we study a twist-driven buckling and fast flip dynamics of a geometrically constraint ribbon by combining experiments, numerical simulations, and analytical theory. We identify two distinct types of shape transitions; a narrow ribbon snaps, whereas a wide ribbon forms a pair of localized helices. We construct a phase diagram and explain the origin of the boundary, which is determined only by geometry. We quantify effects of gravity and clarify time scale dictating the rapid flipping. Our study reveals the unique role of geometric twist-bend coupling on the fast dynamics of a thin constrained structure, which has implications for a wide range of biophysical and applied physical problems. 
\end{abstract}
\maketitle

Transforming elastic potential energy to kinetic energy is a key process in a range of biological and man-made mobile systems. 
A few elegant examples are provided by plants and fungi, which have evolved ingenious mechanisms with elastic instabilities, cavitation, and fracture~\cite{Dumais-AnnuRevFluidMech-2012, Forterre_Nature_2005, Noblin:2012ie, Armon-Science-2011, Hofhuis:2016he}. 
The same physical principles are often found in man-made applications including ancient catapults, children's toys, smart materials and robotics~\cite{Gordon-book, Holms-AdvMat-2007,Trivedi_review_2008,Steltz:2009cu,Pandey_2014_EPL,Bigoni,Jaeger:2015bn,Rus:2015eq,Reis:2015hb, Reis_EML_2015,Yuk:2017gw,Sano:2018ji,Gomez_NatPhys_2016,Yu:2018kn}. 

\begin{figure}[hb!]
\begin{center}
\includegraphics[scale = 0.425]{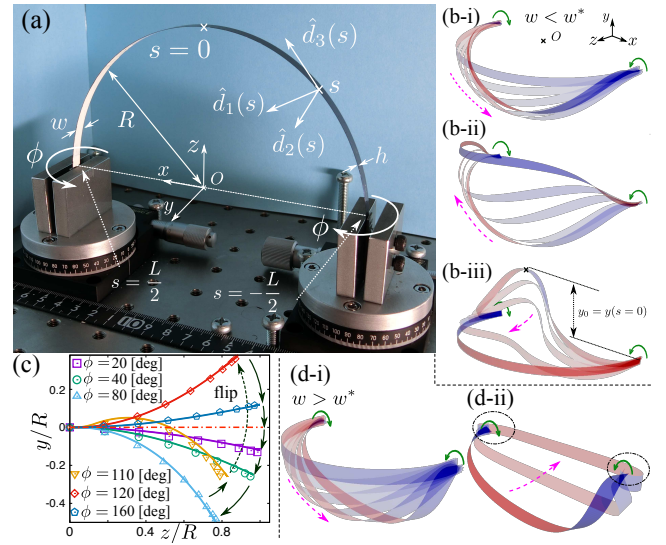}
\caption{(a)~Geometry of experiments and definition of our coordinate system. 
Two ends of a ribbon ($s=\pm L/2$) are clamped with controlled twisting angles $\phi$. 
(b-i)-(b-iii)~Stroboscopic figures of a flipping ribbon obtained from our numerical simulation for $(h,w,R) = (0.2, 8, 115)$~mm. 
Head of the ribbon (b-i) goes down, (b-ii) goes up, and then (b-iii) the ribbon flips. 
(c) Side-view of ribbon centerlines obtained in our simulation (solid lines) and experiment (data points) for various $\phi$ and $(h,w,R) = (0.2, 8, 108)$~mm. 
The chain line, $y = 0$, represents a shape without twist: $\phi = 0, 180^{\circ}$~deg. 
Solid arrows represent the sequence in the experiments, while the dashed line represents the flipping transition. 
(d-i)(d-ii)~Stroboscopic figures of a folding ribbon observed in our simulation for $(h,w,R) = (0.2, 15, 115)$~mm. 
Its head (d-i) goes down and (d-ii) raises to the back. The ribbon never flips in this case. 
}
\label{fig:setup}
\end{center}
\end{figure}

Plant movements often require active processes such as swelling, growth, or stimuli-sensing~\cite{Dumais-AnnuRevFluidMech-2012}.
Contrary to this, we propose a boundary-driven, purely mechanical snapping system that utilizes elastic instabilities of a geometrically constrained slender object. 
An illustration of this phenomenon is easy; Take a strip of paper, hold the two ends so that it forms arc (Fig.~\ref{fig:setup}). 
As one rotates the two ends in the {\it same} direction, the ribbon initially deflects from its original plane, subsequently recoils and eventually flips with a snap to get back its previous configuration. See Supplementary Materials for the movie (S1)~\cite{ref:SM}. 
This simple process with an audible flip provides an avenue to achieve a cyclic snapping motion requiring no additional recovery or active processes, which is a distinct advantage for numerous applications~\cite{Yamada:2010ky,Fukamachi2015PalmtopJA,Sugiyama:2016fq}.

In this paper, we uncover the mechanism by which the elastic energy of our ribbon is stored and rapidly released by combining experiments, numerical simulations and theory. We show that a narrow ribbon snaps, whereas a wide ribbon forms a pair of localized helices, and identify a geometric criterion for snapping to occur. 
The importance of geometric twist-bend coupling has previously been recognized in the study of supercoiling instabilities in biological polymers and fibers~\cite{Moroz-Macromol-1998, Goldstein-PRL-2000} and the rich morphologies of twisted ribbon~\cite{Chopin:2013cq,Chopin:2014tq,Starostin:2007hv,Chopin:2016js,Dias:2014hp,Shen:2015ek}.
Our present study generalizes this result~\cite{vanderHeijden-PhysicaD-1998}, and shows that it also dictates the statics and dynamics of a slender structure.

During the experiments, an intrinsically flat metal ribbon made of spring steel with a Young modulus $E\simeq200$~GPa, Poisson ratio $\nu\simeq0.3$, and mass density $\rho\simeq7.9$ g/cm$^3$ is mounted in a mechanical system forming a semi-circle of radius $R$.
See Fig.~\ref{fig:setup} (a).
The two ends are clamped on the rotary optical stages that control the induced rotational angle $\phi$.
The geometric parameters that characterize a ribbon configuration are the width $w$, thickness $h$, and radius $R$ (related to the arc length $L$ as $L=\pi R$), which are varied in the range 6-16 mm, 0.2-0.4 mm, and 65-178 mm, respectively.
We gradually increase $\phi$ by carefully rotating the optical stages by hand with an accuracy of $1^{\circ}$~deg. 
At each step, the ribbon is observed to reach its equilibrium immediately (except in the direct vicinity of the transition).
The morphologies are recorded with a digital camera, and the resulting images are analyzed further below.
A stiff metal ribbon hardly deforms by its own weight, and its behavior can be effectively understood by entirely neglecting
gravitational forces. 
The effects of gravity for more flexible ribbons will be discussed later.


For a small $\phi$, a ribbon behaves similarly independent of its geometry.
First, it deflects from the original plane of bending, symmetrically developing left- and right-handed helices on each side.
With the increasing $\phi$, we observe two distinctly different behaviors depending on the ribbon's aspect ratio, $w/h$.
A relatively narrow ribbon recoils and eventually flips with a snap, returning to its original configuration but now {\it inside out}.
Because the helices at the two sides have the opposite handedness, they annihilate at the center when they progress, triggering the flip with a snap.    
In contrast, a sufficiently wide ribbon develops localized helices formed at the proximity of the two ends, which remain separated and are stabilized for increasing $\phi$.
At self-contact, a further rotation requires the application of an indefinitely large torque, leading to the formation of creases or kinks. 
We hereafter refer to this mode as ``fold", and the former as ``flip."

To understand the shape bifurcation, we now develop a simple scaling argument for the folding condition based on the energetic balance between bending and stretching~\cite{audoly-book}.
Assuming a helicoidal twisting with twist density per length $\tau$ that scales as $\tau\sim 1/R$, the Gauss-curvature in the ribbon mid-surface is $\sim \tau^2$.
This requires an in-plane strain $\epsilon\sim w^2\tau^2$, leading to a stretching energy given by $E_{\rm str} \sim Ehw^5 R\tau^4$, which should become dominant at the flip transition.
In contrast, when a ribbon folds into helices, its centerline curvature is of the order of $1/R$, and the bending energy may scale as 
$E_{\rm bend} \sim Eh^3w/R$.
The fold configuration is thus realized when $E_{\rm bend}<E_{\rm str}$, which gives the condition
\begin{eqnarray}
w > w^*\sim\sqrt{Rh}
\label{eq:critical_w}.
\end{eqnarray} 
The critical width $w^*$ is independent of the elastic modulus and increases with the induced radius of curvature $R$.
The present argument is thus analogous to the one previously developed for helicoid-spiral transitions in self-assembled chiral-ribbons~\cite{Ghafouri:2005ef, Armon-Softmat-2014,Grossman:2016hi}, in which spontaneous curvatures induce the transition.
The flip-fold transition however gives rise to an intriguing shape dynamics that is absent in the helicoid-spiral ones (as detailed later).

\begin{figure}[!b]
\begin{center}
\includegraphics[scale = 0.95]{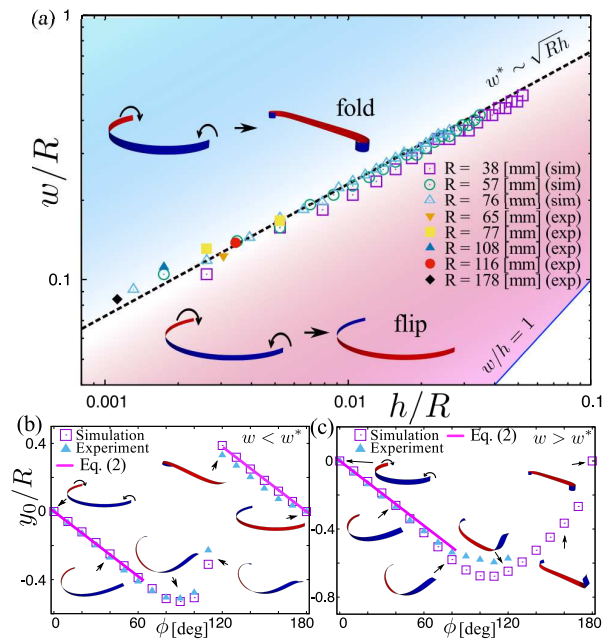}
\caption{(a) Transition-type diagram constructed from our experimental and numerical data for various combination of $(h,w,R)$. 
Dashed line represents the scaling prediction of Eq.~(\ref{eq:critical_w}): $w^*/R = 2.3\sqrt{h/R}$. 
(b)(c) Rescaled center position of a ribbon, $y_0/R$, plotted as a function of angle $\phi$, for (b) narrow ribbon with $(h,w,R) = (0.2, 8, 108)$~mm, and for (c) wide ribbon with $(h,w,R) = (0.2, 15, 108)$~mm.
Open squares and filled triangles represent simulation and experimental data, respectively.
Snapshots are from our simulations. 
Solid lines show our analytical prediction Eq.~(\ref{eq:linear_response_Euler}).}
\label{fig:boundary}
\end{center}
\end{figure}

To complement the experimental data and corroborate the scaling prediction of Eq.~(\ref{eq:critical_w}), we perform numerical simulations based on the Kirchhoff elastic rod formulation for a narrow ribbon, where a weak in-plane stretch is taken into account~\cite{Ghafouri:2005ef}. 
The detailed procedure has previously been described elsewhere~\cite{discrete_model}, and the validity of the method for ribbon mechanics has also been demonstrated recently~\cite{morigaki_prl_2016}. 
All the relevant parameters are matched with the experiments, and there are no adjustable parameters in our simulations (except for the one for the damping forces).

In the simulations, we successfully reproduce the flip and fold transition, where the resulting shapes agree remarkably well with the experimental data~(Fig.~\ref{fig:boundary} (b) and (c)). 
Integrating all the numerical and experimental data, we construct the transition-type diagram in Fig.~\ref{fig:boundary} (a), which verifies our scaling prediction from Eq.~(\ref{eq:critical_w}).
In the remainder of this paper, we will focus on the physics of flipping, for which an analytical descriptions based on the Kirchhoff elastic rod model will apply since the transition is relevant to narrow ribbons satisfying $w/R <(h/R)^{1/2}\ll 1$ (as obtained from Eq.~(\ref{eq:critical_w})).

To quantify the three-dimensional shape changes, we track the position of a specific point on the ribbon centerline at $s=0$ (see Figs.~\ref{fig:setup}(b-iii)) and plotted its $y$-component, $y_0 \equiv y(s = 0)$ in Fig.~\ref{fig:boundary}(b). 
Initially, $y_0$ decreases linearly with $\phi$ and subsequently increases steeply just before the flip. 
(For a wide ribbon with $w>w^*$, $y_0$ also decreases initially, but stays negative throughout the process (Fig.~\ref{fig:boundary}(c))). The linear response can be understood analytically.
To describe the ribbon configuration, we assign at each point of the centerline $s$ an orthogonal director frame $(\hat{\bm d}_1, \hat{\bm d}_2, \hat{\bm d}_3)$, where $\hat{\bm d}_3=\hat{\bm r}'$ is the unit tangent vector ($'$ denotes the derivative with respect to $s$), and $\hat{\bm d}_1$ and $\hat{\bm d}_2$ point towards the principal axes in the ribbon's cross sectional plane~\cite{audoly-book}. 
Assuming the inextensibility, the configuration is determined by specifying how the frame rotates as it moves along the centerline per unit length: 
$\hat{\bm d}_a ' = \bm{\Omega}\times\hat{\bm d}_a$ ($a = 1,2,3$), where the direction of $\bm{\Omega}=\Omega_a\hat{\bm d}_a$ sets the rotational axis at $s$, and $\Omega_a$ gives the rate of rotation, i.e., curvature, about ${\bm d}_a$.
In linear elasticity theory, the elastic deformation energy may be given by
$E_{\rm el} = \int ds (A_1\Omega_1 ^2 + A_2\Omega_2 ^2 + C\Omega_3 ^2)/2$ with the bending and twisting moduli given by $A_1 = Ehw^3/12$, $A_2 = Eh^3w/12$, and $C = Eh^3w/6(1+\nu)$. 
For our present purpose, it is useful to express $\Omega_a$ in terms of the Euler angles, $(\varphi, \theta, \psi)$, where $\varphi$, $\theta$ and $\psi$ represent, an azimuthal, polar, and twist angles, respectively. This leads to $\Omega_1 = \varphi'\cos\theta\sin\psi - \theta'\cos\psi$, $\Omega_2 = \varphi'\cos\theta\cos\psi + \theta'\sin\psi,$ and $\Omega_3 = \varphi'\sin\theta + \psi'$.
To investigate the linear response, we expand the Euler angles around the base state of the semi-circle: $\theta=\psi=0$ and $\varphi = \pi + s/R$.
Retaining the terms up to the quadratic orders in $E_{\rm el}$ and taking the flat-ribbon limit $h/w\to0$, we arrive at the Euler-Lagrange equations for $\psi$ and $\theta$ as ${\theta}' = \psi/R$ and ${\psi}'''' - (\Gamma -2){\psi}''/R^2 + \psi/R^4 = 0$, where $\Gamma = (1+\nu)/2$. 
The equations can be analytically solved for a general $\Gamma$, but the solutions are particularly compact for $\Gamma=0$, where we have
$\psi = - (2\phi/\pi)[(s/R)\sin(s/R) -\cos(s/R)]$. 
The external torque that must be applied to keep the rotational angle $\phi$ at a given value is
$T(\phi)=-C\Omega_3(s = -L/2;\phi) = -4C\phi/\pi R$.
Furthermore, by integrating $\bm{r}'(s) = \hat{\bm d}_3$, one readily determines the centerline shape, and we arrive at 
\begin{eqnarray}
y_0\simeq -\left(1-\frac{2}{\pi}\right)\phi R,
\label{eq:linear_response_Euler}
\end{eqnarray}
which agrees remarkably well with our experimental and numerical results (solid lines in Figs.~\ref{fig:boundary}(b) and (c)) with no fitting parameters. 
Equation~(\ref{eq:linear_response_Euler}) shows that the out-of-plane deflection $y_0$ arises due to coupling between the induced curvature $1/R$ and the rotation (or twist) $\phi$.
The agreement also suggests that the initial response of the ribbon is quite insensitive to the Poisson effect (i.e., the precise value of $\Gamma$), which justify our previous simplifying assumptions. 
Note that Eq.~(\ref{eq:linear_response_Euler}) is also valid for the ribbon shape after the flip, with the replacement $\phi\rightarrow \phi+\pi/2$, as shown in Fig.~\ref{fig:boundary} (b). 

\begin{figure}[!h]
\begin{center}
\includegraphics[scale = 1.7]{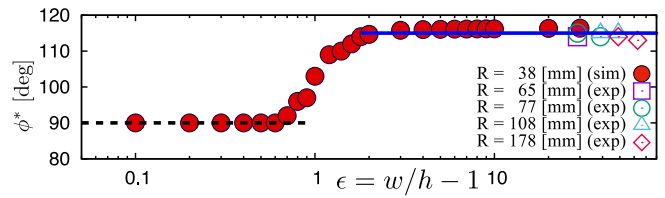}
\caption{Transition angle $\phi^*$ as a function of aspect ratio $\epsilon$ obtained in simulation and experiment for ribbons with $h = 0.2$[mm] and various $R$. The dashed and solid lines denote $90^{\circ}$ and $115^{\circ}$ deg, respectively. 
}
\label{fig:trans_angle_dynamics}
\end{center}
\end{figure}

In between these linear regimes, there is the regime where the nonlinearities become significant. 
Close to the transition, the external torque $T(\phi)$ deviates from the linear relation and rapidly approaches zero, which is a clear sign of the instability. 
This behavior can be most effectively described by expanding $T(\phi)$ as a function of $\phi$, such that
$T(\phi)= -4 \pi^{-1} C R\phi+\gamma \phi^3+\cdots$, where $\gamma$ is assumed to be a positive constant to be determined. 
The critical angle of the flip is known from the condition $T(\phi^*)=0$, which gives $\phi^*=(4 \pi^{-1} C R/\gamma)^{1/2}$.
In Fig.~\ref{fig:trans_angle_dynamics}, $\phi^*$ obtained from our experiments and numerical simulations are plotted as a function of the cross sectional parameter $\epsilon\equiv w/h-1$ (for a circular rod, $\epsilon=0$, while for an infinitely thin ribbon, $\epsilon=\infty$).
The data shows that, while $\phi^*$ is $90^{\circ}$ for $\epsilon\ll 1$, $\phi^*$ approaches $115^{\circ}$ for $\epsilon \rightarrow \infty$~\footnote{Note that an ideally circular rod ($\epsilon=0$) is a singular limit because it would simply rotate axially without any transition.
Note also that the flipping transition for $\epsilon<1$ seems to be continuous, while that for $\epsilon \gg1$ is discontinuous (snap-like).
A detailed study regarding this distinction of the types of transition will be reported elsewhere.}.
This critical angle value is independent of the initial radius $R$ as well as Young's modulus $E$, and is expected to be universal in stiff ribbons for which gravity is negligible.
This independence also suggests the scaling relation $\gamma \sim CR/\phi^{*2}$, i.e., $\gamma$ is directly proportional to $CR$. 
We leave the theoretical explanation of this finding to future work and turn our attention to the small $\epsilon$ regime, for which the centerline curvatures can be calculated analytically beyond the linear response regime. 
Physically, we observe that $\Omega_1\sim \Omega_2 \sim 1/R$, and $\Omega_3 \sim O(\epsilon)$. 
The Kirchhoff rod equations are reduced to a set of the three independent linear ODEs:
$\Omega_1 ' = \Omega_2 ' = 0$ and $\Omega_3 ' = \epsilon(1+\nu)\Omega_1\Omega_2$.
Considering the symmetry of our problem, $\Omega_3\propto s$. 
We then construct the solutions for $\Omega_a$, so that the ribbon shape, which is known by integrating the kinematic relations $\hat{\bm d}_a'= \bm{\Omega}\times\hat{\bm d}_a$, satisfies the boundary conditions at $s = \pm L/2$ and the constraint $\bm{r}(L/2)-\bm{r}(-L/2) = 2R\hat{\bm x}$, which leads to $\Omega_1(\phi) =\sin\phi/R + O(\epsilon)$, $\Omega_2(\phi) =\cos\phi/R+O(\epsilon)$, and 
\begin{equation}
 \Omega_3(s;\phi) = \frac{\epsilon(1+\nu)}{R^2}\frac{\sin2\phi}{2}s\!+ O(\epsilon^2)\label{eq:O3}.
\end{equation}
Thus, the critical angle $\phi^*$ can be derived from $T(\phi^*) =0$ as $\phi^* = \pi/2 + O(\epsilon)$.
Figure~\ref{fig:trans_angle_dynamics} shows that $\phi^*$ starts deviating considerably from $\pi/2$ at $\epsilon\sim 1$ where the higher order terms become significant.


\begin{figure}[h]
\begin{center}
\includegraphics[scale = 0.9]{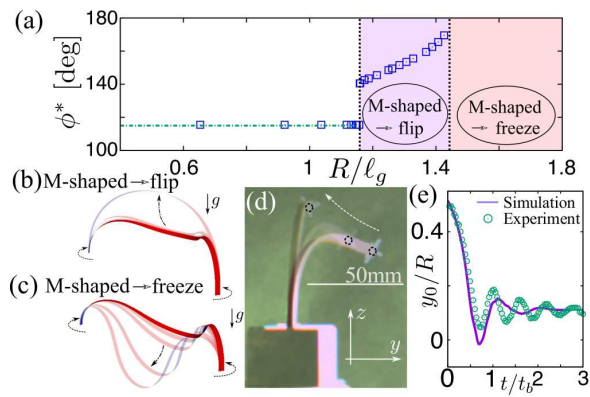}
\caption{(a) $\phi^*$ plotted as a function of $R/\ell_g$ obtained from the numerical simulations of ribbons with $h = 1.0$ mm, $L=240$ mm, and $\nu = 0.5$. In the purple-colored region, the ribbon flips via the M-shape configuration, while flipping is no longer observed in the red region. Stroboscopic figures of (b) a ribbon during flipping from a stable M-shape configuration ($R/\ell_g\simeq1.2$), and of (c) a ribbon that gets stuck in a M-shape configuration ($R/\ell_g\simeq1.5$). (d)(e) The flipping dynamics of elastomer ribbons with $(h,w,R) = (1,4,76)$ mm. (d) Stroboscopic photo of a ribbon at the onset of flipping transition. Circles with broken lines mark the positions of the center of ribbon ($s = 0$) at different times. (e) Comparison between experiment and simulation. Rescaled displacement of the central point, $y_0/R$, plotted as a function of the rescaled time $t/t_b$.}
\label{fig:g_dynamics}
\end{center}
\end{figure}
 
When gravity dominates bending elasticity, the above results change significantly. 
The relative magnitude of gravity over elasticity is quantified by the dimensionless number $R/\ell_g$, where
$\ell_g \sim (Eh^2/\rho g)^{1/3}$ is the gravito-bending length~\cite{wang_review_1986}.
In the limit of stiff rods or vanishing gravity, $R/\ell_g\rightarrow 0$.
For the previously described metal ribbons, we estimate $\ell_g > 200$ mm, which gives $R/\ell_g=0.1-0.5$, thereby justifying our earlier assumption.
Here, instead of the metals, we test a ribbon made of elastomer, for which we can realize the opposite regime characterized by $R/\ell_g >1$.
In our particular system, not only the relative magnitude of gravity but also its relative {\it direction} matters.
If an initial bending plane is horizontal, gravity may act to impede flipping when it first moves downwards upon twisting. 
In the same overall configuration but when the ribbon initially moves upwards (i.e., by twisting it oppositely), gravity may work to assist flipping.
The effects of gravity in these two cases are thus physically easy to understand, and the most non-trivial behavior can be seen when an initial bending plane is set vertical, which we will now investigate in detail.
Figure~\ref{fig:g_dynamics}(a) shows $\phi^*$ obtained from the simulations as a function of $R/\ell_g$. 
For $R/\ell_g<1.16$, $\phi^*\approx 115^{\circ}$ even in presence of gravity, again confirming the validity of our preceding analysis where $R/\ell_g \rightarrow 0$ was assumed.
A novel morphology, termed the "M-shape", appears for $R/\ell_g>1.16$, at a stage prior to flipping (Fig.~\ref{fig:g_dynamics}(b)). 
The ribbon's own weight stabilizes this concave shape, and an additional twisting is necessary to drive the system towards flipping instability~\cite{ref:SM}.
Remarkably, $\phi^*$ increases discontinuously at $R/\ell_g \approx 1.16$, suggesting that the nature of the transition alters from super-critical to sub-critical in the presence of gravity. 
At approximately $R/\ell_g\approx 1.43$, the M-shape is mechanically stabilized further and the flipping from this structure is no longer accessible even for $\phi=180^{\circ}$ (Fig.~\ref{fig:g_dynamics}(c)). 
Thus, the snap-like fast response should be explored for relatively narrow, and stiff ribbons satisfying $R/\ell_g <1$.

To investigate the fast dynamics, we used a high-speed camera (Ditect, HAS-D71, 500 fps) to capture the shape changes. 
For a metal ribbon, we found a rather peculiar, inertial-dominated dynamics~\cite{ref:SM}. 
In Fig.~\ref{fig:g_dynamics}(d) and (e), we instead show the stroboscopic figure and the time evolution of $y_0/R$ onset of flipping obtained from the experiment using an Olefin-based elastomer with its geometric and mechanical parameters being $(h,w,R) = (1,4,76)$ mm, $E\approx 20$ MPa, $\rho\simeq1.1$ g/cm$^3$, and $R/\ell_g\approx 1.3$. 
The flipping transition occurs at $\phi^*\simeq150^{\circ}$ deg. 
The video manifests that large-amplitude bending waves are excited during the shape transition, whose characteristic time is determined by the growth rate of the most unstable mode at the instability.
The linearized equation for an out-of-plane deflection $u$ (in an appropriately defined coordinate) has a well-defined time scale $t_b = R^2/hv_s$, where $v_s$ is the speed of sound $v_s\equiv\sqrt{E/\rho}$~\cite{Audoly:2005cp, Pandey_2014_EPL}.
This arises as the frequency of the flexural wave~\cite{Audoly:2005cp}, but also sets the growth timescale of unstable modes.
In Fig~\ref{fig:g_dynamics}(e), the numerically and experimentally obtained data for $y_0$ are plotted as a function of the rescaled time $t/t_b$, demonstrating that the first period is completed in exactly at $t \sim t_b$. 
The ringing, which could be the origin of the snapping sound, is observed in the experiments at $t>t_b$, but not in the simulations. 
This discrepancy could be resolved by elaborating on our modeling of dissipative processes~\footnote{For the sake of simplicity, a local isotropic drag force proportional to the local velocity in the laboratory (rest) frame is assumed in the simulation. We are improving this by taking internal dissipation within elastomeric materials into account such that the viscous internal stress is proportional to the temporal changes of curvatures of the ribbon's centerline. Our preliminary simulations based on a Kelvin-Voigt viscoelastic ribbon model reproduced the damped oscillations observed in the experiment~\cite{sano_wada_fullpaper}.}.

In this paper, we have investigated the qualitatively new snapping instability of a constrained ribbon, by highlighting the importance of geometric twist-bend coupling.
The results presented are generic, and might be applied to a number of problems in different scales, from torsionally-driven instability of DNA in nucleosomes~\cite{Lavelle-Review-2007, Kouzine-NatSturctBio-2013} to the design of non-muscular engines in soft robotics. 
It is an intriguing open question how the nature of the transition revealed here will be modified by thermal fluctuations, spontaneous curvatures, and/or active processes.

\begin{acknowledgments}
We acknowledge financial support in the form of Grants-in-Aid for Japan Society for the Promotion of Science (JSPS) Fellows (Grant No.~16J05315) and JSPS KAKENHI (Grants No.~15H03712, No.~16H00815, No.~18K13519 and ``Synergy of Fluctuation and Structure: Quest for Universal Laws in Non-Equilibrium Systems").
\end{acknowledgments}

\bibliography{bib_insideout}

\end{document}